\documentclass{article}[11pt]

\newcommand{\ph}[1]{\medbreak \noindent {\bf #1}}

\usepackage{graphicx} 

\title{A Case for Data Commons\protect\\
{\Large Towards Data Science as a Service}}

\author{Robert L. Grossman \and Allison Heath \and Mark Murphy \and Maria Patterson
\protect\\
University of Chicago
\protect\\
\protect\\
Walt Wells
\protect\\
Open Commons Consortium
}
\date{January 1, 2016}

\begin{document}
\maketitle

\section*{Abstract}

As the amount of scientific data continues to grow at ever faster rates, the research community is increasingly in need of flexible computational infrastructure that can support the entirety of the data science lifecycle, including long-term data storage, data exploration and discovery services, and compute capabilities to support data analysis and re-analysis, as new data are added and as scientific pipelines are refined.  We describe our experience developing data commons-- interoperable infrastructure that co-locates data, storage, and compute with common analysis tools-- and present several cases studies.  Across these case studies, several common requirements emerge, including the need for persistent digital identifier and metadata services, APIs, data portability, pay for compute capabilities, and data peering agreements between data commons.  Though many challenges, including sustainability and developing appropriate standards remain, interoperable data commons bring us one step closer to effective Data Science as Service for the scientific research community.

\ph{Keywords:} data commons, software as services, science as a service,
data as a service, cloud computing

\section{Introduction}

With the amount of available scientific data far larger than the ability of the research community to analyze it, there is a critical need for new algorithms, software applications, software services, and cyber-infrastructure to support data throughout its lifecycle in data science.  In this paper, we make a case for the role of data commons in meeting this need. We describe the design and architecture of several data commons that we have developed and operated for the research community in conjunction with the Open Science Data Cloud (OSDC), a multi-petabyte science cloud that the not-for-profit Open Commons Consortium (OCC) has managed and operated since 2009 \cite{Grossman:SC12}.

One of the distinguishing characteristics of the OSDC is that it interoperates with a data commons containing over 1 PB of public research data through a service-based architecture.  This is an example of what is sometimes called ``Data as a Service,'' which play an important role in some Science as as Service frameworks.

There are at least two definitions for {\em Science as a Service}.  The first is analogous to the Software as a Service \cite{Mell:2011} model in which instead of managing data and software locally using your own storage and computing resources, one uses the storage, computing, and software services offered by a Cloud Service Provider (CSP).  With this approach, instead of setting up your own storage and computing infrastructure and installing the required software, a scientist uploads data to a CSP and uses pre-installed software for data analysis.  Note that a trained scientist is still required to run the software and analyze the data.

Science as a Service can also refer more generally to a service model that relaxes the requirement of needing a trained scientist to process and analyze the data.  With this service model, specific software and analysis tools are available for specific types of scientific data.  Data are uploaded to the Science as a Service Provider, processed using the appropriate pipelines, and then made available to the researcher for further analysis if required.   

Obviously these two definitions are closely connected in that a scientist can set up the required Science as a Service framework as in the first definition so that less trained technicians can use the service to process their research data as in the second definition.  By and large, we focus on the first definition in this paper.

We close this section by discussing Science as a Service more generally and how data commons fit in.
There are various Science as a Service frameworks, including variants of the types of clouds formalized by NIST in \cite{Mell:2011} (Infrastructure as a Service, Platform as a Service, and Software as a Service), as well as some more specialized services that are relevant for data science (data science support services and data commons).  These Science as a Service frameworks include the following:

\begin{itemize}

\item Data science infrastructure and platform services, in which virtual machines, containers, or platform environments containing commonly used applications, tools, services, and datasets are made available to researchers. The OSDC is an example of this service model \cite{Grossman:SC12}.

\item Data science Software as a Service, in which data are uploaded and processed by one or more pipelines, with the results being downloaded or stored in a cloud. There are general purpose platforms offering data science as a service, such as Agave \cite{dooley2012agave}, as well as more specialized services, such as those designed to process genomics data.  For example, there are several commercial companies that enable a researcher to upload genomic sequence data, select a bioinformatics pipeline, and download the processed data.  

\item Data science support services, including data storage services, data sharing services, data transfer services, data collaboration services, etc.  A prominent example of this model is Globus \cite{foster2011globus}.

\item Data commons, in which data, data science computing infrastructure, and data science applications and services are co-located.  We describe data commons in more detail in Section~\ref{section:data-commons} below.

\end{itemize}

The paper is organized as follows.  In Section~\ref{data-commons-section}, we introduce data commons. In Section~\ref{section:osdc}, we introduce the Open Science Data Cloud and some of the data commons associated with it.  In Section~\ref{section:design}, we describe the design and architecture of the OSDC, which includes the services required to set up, manage, and operate data commons.  Section~\ref{section:case-studies} contains two case studies of OSDC data commons, and Section~\ref{section:discussion} contains some challenges related to data commons.

\section{Data Commons}
\label{data-commons-section}

\subsection{What is a Data Commons?}
\label{section:data-commons}

By a {\em data commons} we mean cyber-infrastructure that co-locates data, storage, and computing infrastructure with commonly used tools for analyzing and sharing data to create an interoperable resource for the research community.

In the discussion below, we distinguish between several stakeholders that are involved in data commons: 1) the {\em data commons service provider (DCSP)}, the entity operating the data commons; 2) the {\em data contributor (DC)}, the individual or organization providing the data to the DCSP; and 3) the {\em data user (DU)}, the researcher or organization that accesses the data.  Note that often there is a fourth stakeholder: the DCSP associated with the researcher accessing the data.  In general, there will be an agreement, often called the Data Contributors Agreement (DCA), governing the terms by which the data is managed by the DCSP and the researchers accessing the data, and an agreement, often called the Data Access Agreement (DAA), governing the terms of any researcher that accesses the data.

As we describe in more detail below, we have built several data commons since 2009.  Based upon this experience we have identified six main requirements that, if followed, would enable data commons to interoperate with other data commons, with science clouds \cite{Grossman:SC12}, and with other cyber-infrastructure supporting Science as a Service.

\ph{Requirement 1: Permanent Digital IDs.} 
The first requirement is that the data commons must have a digital ID service and that datasets in the data commons must have permanent, persistent digital IDs.  Digital IDs are described in more detail in Section~\ref{section:core-services}.  Associated with digital IDs are i) access controls specifying who can access the data and ii) metadata specifying additional information about the data (see Requirement 2).   Part of this requirement is that data can be accessed from the data commons through an API by specifying its digital ID.

\ph{Requirement 2: Permanent Metadata.}
The second requirement is that there is a metadata service that for each digital ID returns the associated metadata.  Since the metadata can be indexed, this provides a basic mechanism for the data to be discoverable. 

\ph{Requirement 3: API-based Access.}
The third requirement is that data can be accessed by an application programming interface (API), not just by browsing through a portal. Part of this requirement is that there is a metadata service that can be queried that returns a list of digital IDs that can then be retrieved via the API.  For those data commons that contain controlled access data, another component of the requirement is that there is an authentication and authorization service so that users can first be authenticated and the data commons can check whether they are authorized to have access to the data.

\ph{Requirement 4: Data Portability.}
The fourth main requirement is that data be portable in the sense that a dataset that has been contributed to a data commons can be transported to another data commons so that in the future it is hosted by the second data commons.  In general, if data access is through digital IDs (versus referencing the physical location of data), then software that references data should not have to be changed when data is re-hosted by a second data commons.

\ph{Requirement 5: Data Peering.}  By data peering, we mean an agreement between two data commons service providers to transfer data {\em at no cost} so that a researcher at data commons 1 can access data stored at data commons 2.  In other words, the two data commons agree to transport research data between them with no access charges, no egress charges, and no ingress charges. 

\ph{Requirement 6: Pay for Compute.}  The final requirement co-located with the data commons is computing infrastructure that is available to researchers.  Since, in practice, researchers' demand for computing resources is larger than the available computing resources, computing resources must be rationed, either through allocations or by charging for the use of computing resources.  Notice that there is an asymmetry in how a data commons treats the storage and computing infrastructure. When data are accepted into a data commons, there is a commitment to store the data and make it available for a certain period of time, often indefinitely.   In other words, the rationing decision for initial storage is made when data are accepted.  In contrast, computing over data in a data commons is rationed in an on-going fashion, as is the working storage, and the storage required for derived data products, either by providing computing and storage allocations for this purpose or by charging for computing and storage.  For simplicity, we refer to this requirement as ``pay for computing," even though the model is more complicated than that.

Although very important for many applications, we view other services, such as services providing data provenance \cite{simmhan2005survey}, data replication \cite{chervenak2008wide}, and data collaboration \cite{alameda2007open}, as optional additional services, and not as core services.

\section{Open Science Data Cloud and OCC Data Commons}
\label{section:osdc}

\subsection{About the Open Science Data Cloud and the OCC}
The Open Science Data Cloud (OSDC) is a multi-petabyte science cloud that serves the research community by co-locating a multidisciplinary data commons containing 
approximately 1 PB of scientific data with cloud-based computing, high
performance data transport services, and virtual machine images and shareable snapshots containing common data analysis pipelines and tools.   

The OSDC is designed to provide a long term persistent home for scientific data, as well as a platform for data intensive science allowing new types of data
intensive algorithms to be developed, tested, and used over large sets of heterogeneous scientific data. Recently, OSDC researchers have logged about 2 million 
core hours each month, which translates to over \$800,000 worth of cloud 
computing services if purchased through Amazon's AWS public cloud.
This is over 12,000 core hours per user or, a 16 core machine continuously used by each researcher on average.  

OSDC researchers used a total of over 18 million core hours in 2015.  
We currently target operating OSDC computing resources at approximately 85\% of capacity and storage resources at 80\% of capacity.   Given these constraints, we determine how many researchers to support and what size allocations to provide them.  Since the OSDC specializes in supporting data intensive research projects, we have chosen to target researchers who need larger scale resources (relative to our total capacity) for data intensive science. In other words, rather than support more researchers with smaller allocations, we support fewer researchers with larger allocations.   Table \ref{tab:users} shows the number of times researchers exceeded the indicated number of core hours in a single month during 2015.  

The OSDC is developed and operated by 
the Open Commons Consortium (OCC), which is a 501(c)(3) not-for-profit supporting the 
scientific community by operating data commons and cloud computing infrastructure to 
support scientific, environmental, medical, and healthcare related research.  
OCC Members and Partners include universities (e.g., University of Chicago,
Northwestern University, University of Michigan); companies (e.g., Yahoo!, Cisco,
Infoblox); U.S. government agencies and national laboratories (e.g., NASA, NOAA); 
and international partners (e.g., Edinburgh University, University of
Amsterdam, AIST in Japan).  The OSDC is a joint project with the University of Chicago, which provides the data center used by the OSDC.   Much of the support for
the OSDC came from the Moore Foundation and from corporate donations.

\begin{table}[!ht]
\caption{Data intensive users supported by the OSDC}
\begin{center}
\begin{tabular}{|l | r|}
\hline
\# core hours during month & \# of users \\
\hline
20,000 & 120  \\ \hline
50,000 &  34   \\ \hline
100,000 & 23  \\ \hline
200,000 &  5  \\ 
\hline 
\end{tabular}
\end{center}
\label{tab:users}
\footnotesize{The estimated cost of 100,000 core hours on a commercial cloud service provider like AWS is \$40,000 per month.}
\end{table}

\subsection{The OSDC Community}

The OSDC has a wide-reaching, multi-campus, multi-institutional,
interdisciplinary user base and has supported over 760 research
projects since its inception.  In November 2015, 186
researchers used the OSDC, out of a total of 470 allocation recipients
from 54 universities and research organizations in 14 countries who
were active sometime during 2015.  The most
computational intensive group projects in 2015 included
projects around biological sciences and genomics research, analysis of
earth science satellite imagery data, analysis of text data in historical and scientific
literature, and a computationally intensive project in sociology.  

\subsection{OCC Data Commons}

\ph{OSDC Data Commons.}
We introduced our first data commons in 2009.  It currently 
holds approximately 800 TB of public open access research data, including earth science data, biological data, social science data, and digital humanities data.

\ph{Matsu Data Commons.}
The Open Cloud Consortium has collaborated with NASA since 2009 in
Project Matsu.  Project Matsu's data commons contains 6 years of EO-1
data with new data added daily, as well as selected datasets from other NASA satellites,
including MODIS and the Landsat Global Land Surveys.  

\ph{The OCC NOAA Data Commons.}
In April, 2015, NOAA announced 5 data
alliance partnerships that would have broad access to NOAA's data and
help make these data more accessible to the public.  The NOAA Data Alliance
Partners are Amazon, Google, IBM, Microsoft, and the OCC.  
Currently, only a small fraction of the over 20 PB of data that NOAA has available
in its archives is available to the public.   The NOAA Data Alliance
Partners have broader access to the data in the NOAA repositories.
The focus of the OCC data alliance is to work with the environmental research community to build 
an environmental data commons.   Currently the OCC NOAA Data Commons contains NEXRAD data.
Additional datasets will be added in 2016.

\ph{National Cancer Institute's Genomic Data Commons (GDC).}
Through a contract between NCI and the University of Chicago and in
collaboration with the OCC, we have developed a data commons for
cancer data called the Genomic Data Commons (GDC).   The GDC contains
genomic data and the associated clinical data from NCI
funded projects.  Currently the GDC contains about 2 PB of data, but this is
expected to grow rapidly over the next few years.  The GDC is in beta
testing now (January, 2016) and will be accessible to the public in May, 2016.

\ph{Bionimbus Protected Data Cloud.}
We also operate two private cloud computing platforms that are designed
to hold human genomic data and other sensitive biomedical data.  These
two clouds contain a variety of sensitive controlled access biomedical data that we make
available to the research community following the requirements of
the relevant data access committees.

\ph{Common software stack.}
The core software stack for the various data commons and
clouds described above is all open source.  Many of the components are
developed by third parties, but some key services described below are
developed and maintained by the OCC and other working groups.  Although there are some
differences between them, we try to keep the software stack as close
as possible across the various data commons. 
In practice, as we develop new
versions of the basic software stack, it usually takes a year or so
until the changes can percolate throughout our entire infrastructure.

\section{Design and Architecture}
\label{section:design}

\subsection{OSDC Architecture}
\begin{figure}
\centering
\includegraphics[scale=0.75]{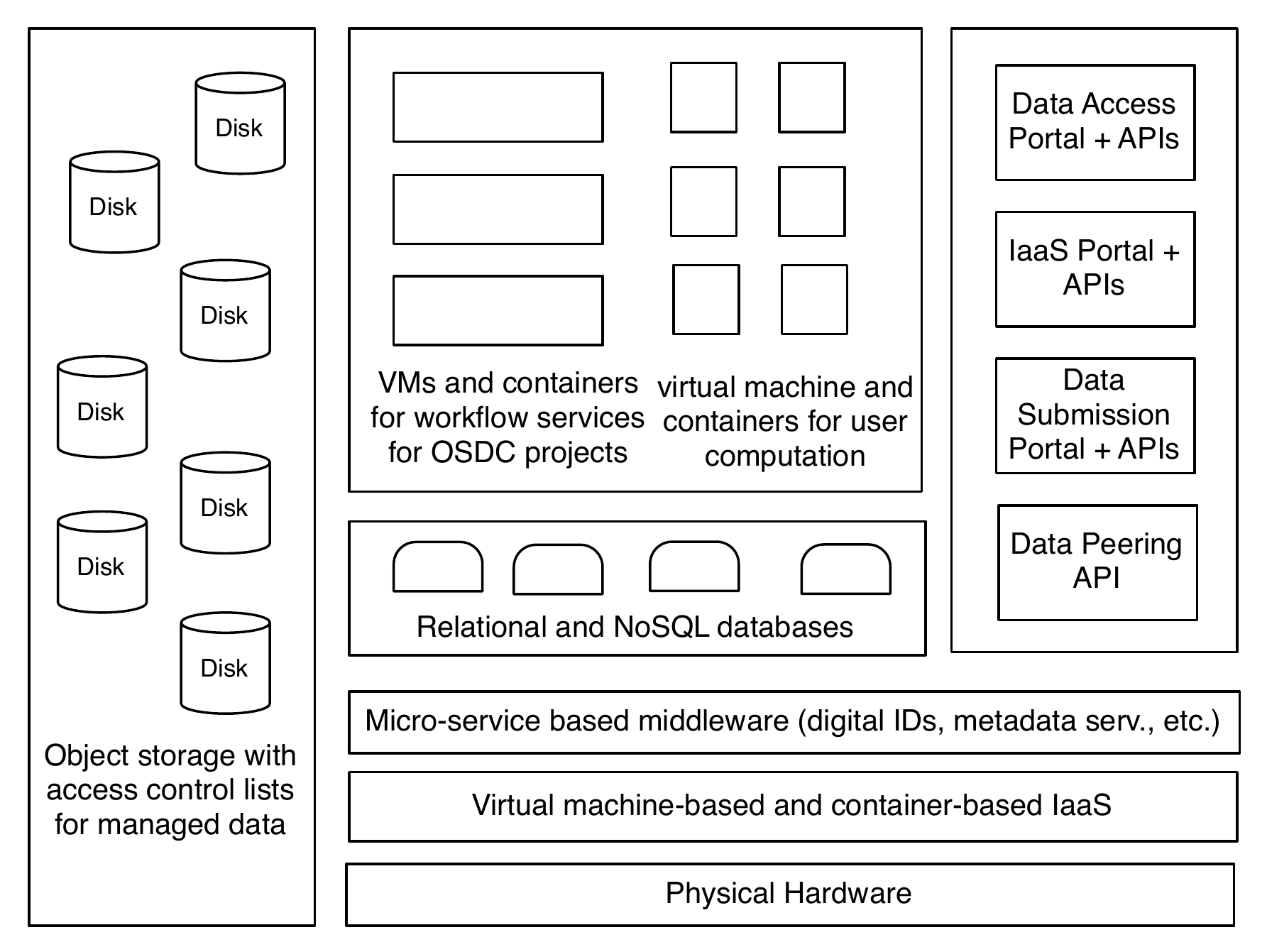}
\caption{The OSDC architecture}
\label{figure:osdc-arch}
\end{figure}

The architecture of the OSDC is shown in Figure~\ref{figure:osdc-arch}.
We are currently transitioning from version 2 of the OSDC software stack \cite{Grossman:SC12} to version
3.  Both versions of the software stack are based upon OpenStack \cite{pepple2011deploying} for Infrastructure as a Service.  Version 2 uses
GlusterFS \cite{davies2013scale} for storage,  while version 3 uses Ceph \cite{weil2006ceph} for block and object storage.  

The OSDC has a portal called the Tukey Portal, which provides a front-end web portal 
interface for users to access, launch, and manage VMs and storage.
The Tukey Portal interfaces with the Tukey Middleware, which provides a secure authentication
layer and interface between various software stacks.
OSDC uses federated login for authentication so that academic institutions with InCommon, 
CANARIE, or the UK Federation can use those credentials. We have worked with 145 
academic universities and research institutions so far to release the appropriate attributes
for authentication.  We also support Gmail and Yahoo logins, but only for approved projects, and when other authentication options are not available.

We instrument all of the resources that we operate so that we can meter and collect the data required for accounting and billing for each user. We use Salesforce.com,
one of the components of the OSDC that is not open source, to send out invoices.
Even when computing resources are allocated and no payment is required, we have found that receipt of these invoices promotes responsible usage of OSDC community resources.
We also operate an interactive support ticketing system that tracks user support
requests and systems team responses for technical questions.
Collecting these data enables us to track usage statistics and to build a 
comprehensive assessment of how researchers use our services.

While adding to our resources, we have developed an infrastructure automation tool called Yates to simplify bringing up new computing, storage, and networking infrastructure. We also try to automate as much of the security required to operate the
OSDC as is practical.


The core OSDC software stack is open source, enabling those interested to set up their own science cloud or data commons.  The core software stack consists of third party open source software, such as OpenStack and Ceph, and open source software developed by the OSDC community.  The latter is licensed under the open source Apache license.  The OSDC does use some proprietary software, such as Salesforce.com to do the accounting and billing, as was mentioned above.


\subsection{OCC Digital ID and Metadata Services}
\label{section:core-services}

In this section, we describe the digital ID service and the associated metadata service used in the data commons.   The digital ID service is accessible via an API that generates digital IDs, assigns
key-value attributes to DIDs, and returns key-value attributes associated with digital IDs.
We have also developed a metadata service that is accessible
via an API and can assign and retrieve metadata associated with a digital ID. 
Users can also edit metadata associated with digital IDs if they have write access to the
digital ID.  These are described in more detail below.  Currently, due to different release schedules, there are some differences in the digital ID and metadata services between several of the data commons that we operate, but over time we plan to converge these services.

\ph{Persistent identifier strategies.} Although the necessity of assigning digital IDs to data is well
recognized  \cite{Mayernik2012data, Duerr2011utility}, there is not yet
a widely accepted service for this purpose, especially for large
datasets \cite{lagoze2014ced}.  This is in contrast to the generally accepted use of digital
object identifiers (DOI) or handles for referencing digital publications.
An alternative to a DOI is an ARK, which
is a Uniform Resource Locator (URL) that is a multi-purpose identifier for information objects of any type
\cite{kunze2003towards, Kunze:2006}.  
In practice, DOIs or ARKs are generally used to assign IDs to datasets, with individual
communities sometimes developing their own IDs. 
%
DataCite is an international consortium that manages DOIs for datasets
and supports services for finding, accessing, and reusing data
\cite{Pollard2010making}.   There are also services such as EZID that
support both DOIs and ARKs \cite{starr2012collaborative}. 

Given the challenges the community is facing coming to a consensus
about which digital IDs to use, our approach has been to build an open
source digital ID service that can support multiple digital IDs and
that can scale to large datasets.  This approach also has the
advantage that we can deal with issues like ``suffix pass-through''
\cite{lagoze2014ced} and associating availability zones \cite{varia2010architecting} and support multiple access methods with the digital ID service that are helpful with petabyte-scale data archives.

\ph{Benefits of digital IDs.} From the viewpoint of researchers, the need for digital IDs associated with datasets is well appreciated \cite{ball2011cite, green2009we}.  Here we discuss some of the reasons that digital IDs are important for a data commons from an operational point of view.  

First, with digital IDs, data can be moved from one physical location or storage system within a data commons to another without the necessity of changing any code that references the data.  As the amount of data grows, moving data between zones within a data commons or between storage systems becomes more and more common and digital IDs allow this to take place without impacting researchers.   

Second, digital IDs are an important component of the data portability requirement.  More specifically, datasets can be moved between data commons, and, again, researchers do not need to change their code.
In practice, datasets can be migrated over time, with the digital IDs references updated as the migration proceeds.

\ph{Signpost digital ID service.}  Signpost is the digital ID service for the OSDC.  
Instead of using a hardcoded URL, the primary way to access managed data 
via the OSDC is through a digital ID. Signpost is an implementation of this concept via JSON documents. 

%

%

The Signpost digital ID service integrates a mutable ID that is assigned to the data along with an immutable hash-based ID that is computed from the data.  Both IDs are accessible through a REST API interface.
With this approach, data contributors can may make updates to the data and retain the same ID, while the data commons service provider can use the hash-based ID to facilitate the management of the data.
In order to prevent unauthorized editing of digital IDs, an access control list (ACL) is kept by each
digital ID specifying the read/write permissions for different users and groups. 

The top layer utilizes \textit{user-defined identifiers}.  These are flexible, may be of any 
format, including Archival Resource Keys (ARKs) and Digital Object Identifiers (DOIs), and 
provide a layer of human readability.  These identifiers map to hashes of the identified
data objects.  
The bottom layer utilizes \textit{hash-based identifiers}.  Hash-based identifiers guarantee 
immutability of the data, allow for identification of duplicated data via hash
collisions, and allow for verification upon retrieval.  These map to known locations of 
the identified data.


Below, we show an implementation of the two-layer identifier service.
A researcher who has a human-readable identifier, for example, provided by a collaborating
researcher or publication or returned by a metadata search service can query Signpost to 
return the requested data.
Alternatively, a researcher who knows the exact hash of a piece of data (e.g., a scientist who is given a data file directly by a collaborator) can also query that way.


\begin{verbatim}
curl http://<signpost>/alias/ark:/31807/DC0-7b2c1002-e3c4-41ea-8edc-8fcee4ff3f47
   {"hashes": { "md5": "1e24480435408b664b756be0822028a3" }, 
    "authority": "noaa-commons", 
    "metadata": <metadata-record-id>, 
    "name": "ark:/31807/DC0-7b2c1002-e3c4-41ea-8edc-8fcee4ff3f47", 
    "release": "public", 
    "rev": "e1407bdd", 
    "size": 45893621760, 
    "urls": [ "https://<osdc>/noaa-nexrad-l2/NWS_NEXRAD_NXL2DP_KDVN_201509_01.tar", 
       "https://<osdc>/noaa-nexrad-l2/NWS_NEXRAD_NXL2DP_KDVN_201509_02.tar", ... ] }
\end{verbatim}

In the next example, a curl request directly to the hash-based layer of the 
identifier service returns a list of urls to the data object.  

\begin{verbatim}
curl http://<signpost>/urls/?hash=md5:1e24480435408b664b756be0822028a3&size=45893621760
    { "hashes": { "md5": "1e24480435408b664b756be0822028a3" }, 
    "size": 45893621760, 
    "urls": [ "https://<osdc>/noaa-nexrad-l2/NWS_NEXRAD_NXL2DP_KDVN_201509_01.tar", 
       "https://<osdc>/noaa-nexrad-l2/NWS_NEXRAD_NXL2DP_KDVN_201509_02.tar", ... ] }
\end{verbatim}

In either case, the data can be moved and the urls updated without affecting any code that 
interfaces with data in the commons using the indexing service.  The concept is simple
and lightweight enough that it can be 
easily used in many existing environments.

\ph{OSDC metadata service.}
The OSDC metadata service is called Sightseer.  Sightseer allows users to
create, modify and access searchable JSON documents containing
metadata about digital IDs.  The primary data can be accessed using
Signpost and the digital ID.
At its core, Sightseer provides no restrictions on the JSON documents
it can store. However, it has the ability to specify metadata
types and associate them with JSON schemas. This helps prevent
unexpected errors in metadata with defined schemes. Sightseer has
similar abilities as Signpost to provide access control lists to
specify users that have write/read access to the specific JSON
document.

\section{Case Studies}
\label{section:case-studies}

\subsection{Matsu}
Project Matsu is an OCC data commons that has been operational since 2009.  
It is a collaboration between NASA and the OCC that is hosted by the University of Chicago, processes the data produced each day by NASA's Earth Observing-1
(EO-1) satellite, and makes a variety of data products available to the research community, including flood maps. The raw data, processed data, and data products are all available through OSDC. Project Matsu uses a framework called the OSDC Wheel to ingest raw data, process
and analyze it, and deliver reports with actionable information to the community in near real time \cite{mandl2013use}.

\begin{figure}[!ht] 
\centering
\includegraphics[scale = .25]{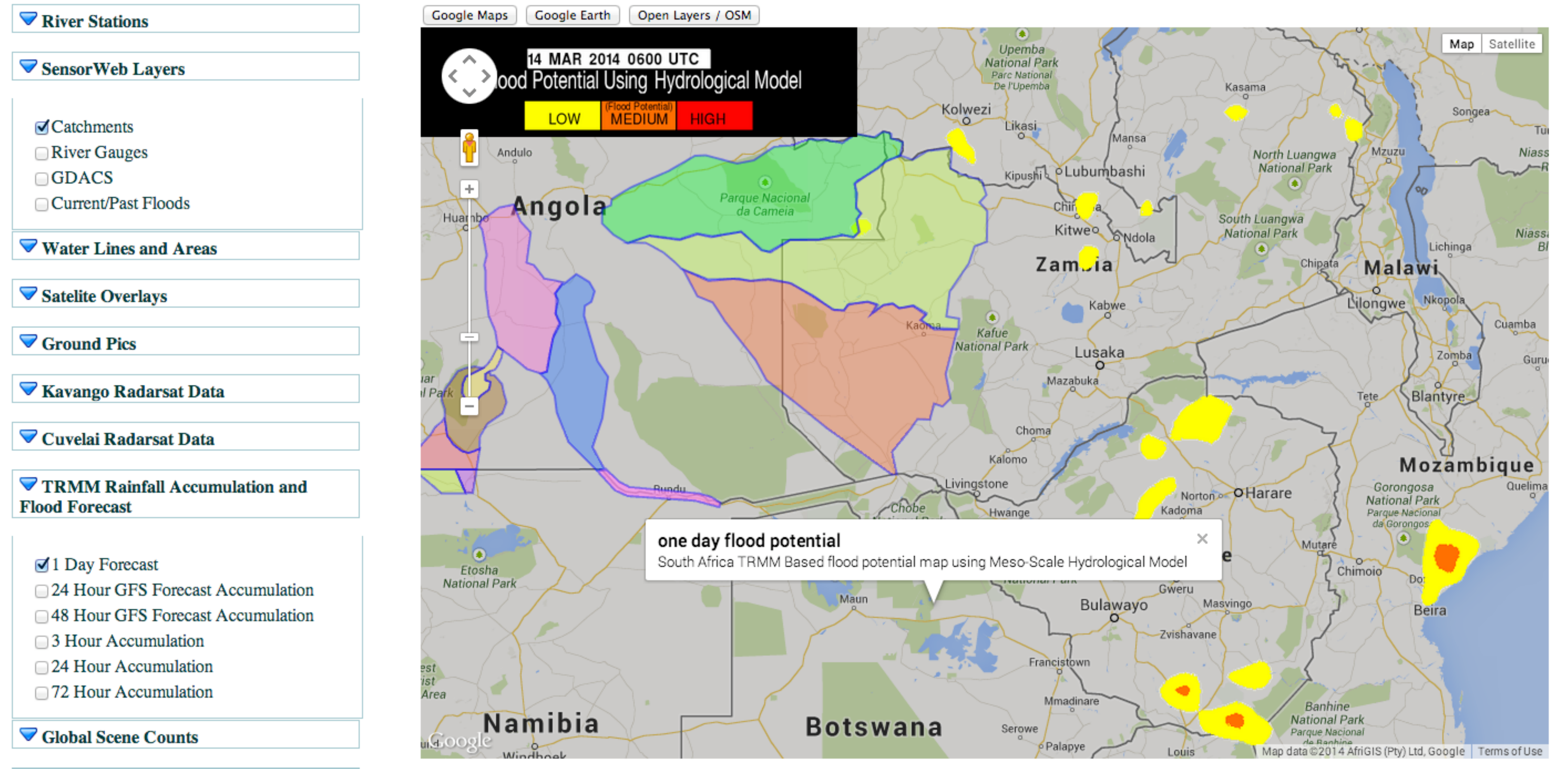}
\caption{\footnotesize A screenshot of part of the Namibia Flood Dashboard from March 
14, 2014 showing water catchments (outlined and colored regions) and a one day flood 
potential forecast of the area from hydrological models using data from the Tropical 
Rainfall Measuring Mission (TRMM), a joint space mission between NASA and the
Japan Aerospace Exploration Agency (JAXA).
}
\label{fig:flood dashboard}
\end{figure}

As part of Project Matsu, we host several focused analytic products with value-added data.
In Figure \ref{fig:flood dashboard} we show a screenshot from one of these focused analytic
products, the Project Matsu Namibia Flood Dashboard \cite{mandl2013use}.  
The Namibia Flood Dashboard was 
developed as a tool for aggregating and rapidly presenting data and sources of information 
about ground conditions, rainfall, and other hydrological information to 
citizens and decision makers in the flood prone areas of water basins in Namibia and 
the surrounding areas.  The tool features a bulletin system producing a short daily written
report, a geospatial data visualization display using Google Maps/Earth and OpenStreetMap, 
methods for retrieving NASA images in a region of interest, and analytics for projecting
flood potential using hydrological models.  The Namibia Flood Dashboard is an important
tool for developing better situational awareness and enabling fast decision making and
is a model for the types of focused analytics products made possible by co-locating 
related datasets with each other and with computational and analytic capabilities.  

\subsection{Bionimbus}

The Bionimbus Protected Data Cloud is a petabyte scale private cloud and data commons that has been operational since March 13, 2013.  Since going online in 2013, Bionimbus has supported more than 152 allocation recipients from over 35 different projects at 29 different institutions. Each month, Bionimbus provides over 2.5 million core hours to researchers, which at standard Amazon AWS pricing would cost over \$500,000. 
One of the largest users of Bionimbus is The Cancer Genome Atlas (TCGA) / International Cancer Genome Consortium (ICGC) PanCancer Analysis of Whole Genomes working group (PCAWG).  PCAWG is currently undertaking a large scale analysis of most of the world's whole genome cancer data available to the cancer community through the TCGA and ICGA consortia using several clouds, including Bionimbus.

Bionimbus is an NIH Trusted Partner \cite{nihGDS16} that interoperates with eRA commons to authenticate researchers and with dbGaP to authorize user’s access to specific controlled access datasets, such as the TCGA dataset.


The design and architecture for Bionimbus is described in \cite{Grossman:JAMIA-14}. The current architecture uses OpenStack for providing virtualized infrastructure, containers for providing a platform as a service capability, and object-based storage with an AWS compatible interface. See Figure~\ref{figure:osdc-arch}.


\section{Discussion}
\label{section:discussion}

\subsection{Data Commons and Science as a Service }


With the appropriate services, data commons support three different, but related, functions.
First, data commons can serve as a data repository or digital library for data that is associated with published research.  Second, data commons can store data, along with computational environments in virtual machines or containers, so that computations supporting scientific discoveries can be reproducible.   Third, data commons can service as a platform enabling future discoveries, as more data, new algorithms, or new software applications are added to the commons.

Data commons fit well with a Science as a Service model: although data commons allow researchers to download data, host it themselves, and analyze it locally, they also support a Science as a Service model, in which i) current data can be reanalyzed with new methods and new tools and applications using co-located computing infrastructures; ii) new data can be uploaded for an integrated analysis; and iii) hosted data can be made available to other resources and applications using a data-as-a-service model in which data in a data commons is accessed through an API.  A data-as-a-service model is enhanced when multiple data commons and science clouds peer so that data may be moved between them at no cost.

\subsection{Challenges}

\ph{Sustainability challenges.}
Perhaps the biggest challenge for data commons, especially large scale data commons, is developing long-term sustainability models that support operations year after year. 

Over the past several years, funding agencies have required data management plans for the dissemination and sharing of research results, but, by and large, have not provided funding to support this requirement.   What this means is that there are a lot of data searching for data commons and similar infrastructure, but very little funding available to support this type of infrastructure.


\ph{Research Challenges.}
Data centers are sometimes divided into several ``pods'' to facilitate their management and build out.  For lack of better name, we sometimes use the term {\em cyberpod} to refer to the scale of a pod at a data center. A pod might contain 50 to several hundred racks of computing infrastructure.  Large scale Internet companies, such as 
Google \cite{dean2010mapreduce} and Amazon \cite{vogels2007dynamo}, have software that scales to cyberpods, but this proprietary software is not available to the research community.  
Although there are software applications, such as Hadoop \cite{white2012hadoop}, that are available to the research community and scale across multiple racks, there is not a complete open source software stack containing all the services required to build a large scale data commons, including the infrastructure automation and management services, security services, etc. \cite{barroso2013datacenter} that are required to operate a data commons at scale. 

We single out three research challenges related to building data commons at the scale of cyberpods.

\medbreak\noindent
{\em Software stacks for cyberpods.} 
The first research challenge is to develop a scalable open source software stack that the provides that infrastructure automation and monitoring, computing, storage, security services and related services required to operate computing services at the scale of a cyberpod \cite{barroso2013datacenter}.

\medbreak\noindent
{\em Datapods.} The second research challenge is to develop data management services that scale out to cyberpods.  We sometimes use the term {\em datapods} for data management infrastructure at this scale.

\medbreak\noindent
{\em AnalyticOps.} The third challenge is to develop an integrated development and operations methodology to support large scale analysis and re-analysis of data.   You might think of this as the analogy of DevOps for large scale data analysis.

\ph{Community challenges.}   The final category of challenges is the lack of consensus within the research community for a core set of standards that would support data commons.  There are not yet widely accepted standards for indexing data, APIs for accessing data, and authentication and authorization protocols for accessing controlled access data.

\subsection{Lessons Learned}

\ph{Data re-analysis is an important capability.}  For many research projects, large datasets are periodically re-analyzed using new algorithms or new software applications.  Data commons are a convenient and cost effective way to provide this service, especially as the size of the data grows and it becomes more expensive to transfer the data.

\ph{Important discoveries are made at all computing resource levels.}  As mentioned, 
computing resources are rationed in a data commons (either directly through allocations or 
indirectly through charge backs).  Typically, there is a range of requests for computing allocations in a data commons spanning 6 to 7 or more orders of magnitude, ranging from hundreds of core hours to tens of millions of core hours.  The challenge is that important discoveries are usually made across the entire range of resource allocations, from the smallest to the largest.   This is because when large datasets, especially multiple large datasets, are co-located it is possible to make interesting discoveries even with relatively small amounts of compute.

\ph{Higher order services.}  Over the past several years, much of the research focus has been designing and operating data commons and science clouds that are scalable, contain interesting datasets, and offer computing infrastructure as a service.   We expect that as these types of Science as a Service offerings become more common, there will be a variety of more interesting higher order services, including discovery, correlation, and other analysis services that are offered within a commons or cloud and across several that interoperate. 

\ph{Cross-commons services.}  
Today, web mashups are quite common, but analysis mashups in which data are left in place but continuously analyzed as a distributed service are relatively rare.  As data commons and science clouds become more common, these types of services can be more easily built.

\ph{Hybrid clouds will become the norm.}
At the scale of a several dozen racks (a cyberpod), a highly utilized data commons in a well-run data center is less expensive than using today's public clouds \cite{Grossman:JAMIA-14}.  For this reason, hybrid clouds consisting of privately run cyberpods hosting data commons that interoperate with public clouds seem to have certain advantages.

\section{Summary and Conclusion}

In this paper, we have described a core set of requirements for data commons and some of our experiences developing and operating data commons with this design as part of the Open Science Data Cloud.  Properly designed data commons can serve several roles in Science as Service:  First, they can serve as an active, accessible, citable repository for research data in general, and research data associated with published research papers in particular.  Second, by co-locating computing resources, they can serve as a platform for reproducing research results.   Third, they can support future discoveries as more data are added to the commons, as new algorithms are developed and implemented in the commons, and as new software applications and tools are integrated into the commons.  Fourth, they can serve as a core component in an interoperable ``web of data,'' as the number of data commons begins to grow, as standards for data commons and their interoperability begin to mature, and as data commons begin to peer.

\section*{Acknowledgments}
This material is based in part upon work supported by the National Science Foundation under Grant Numbers OISE 1129076, CISE 1127316 and CISE 1251201 and by NIH/Leidos Biomedical Research, Inc. through contracts 14X050 and 13XS021 / HHSN261200800001E.

\clearpage
\bibliographystyle{IEEEtran}
\bibliography{bib-15}

\end{document}